\documentclass{pasj00}

\newcommand{{\asc}}{\rm ASCA}
\newcommand{{\suz}}{\rm Suzaku}
\newcommand{{\chan}}{\rm Chandra}
\newcommand{{\dem}}{DEM~L71}
\newcommand{{\psrc}}{J050550.3$-$675017}
\newcommand{{\ka}}{K$\alpha$}
\newcommand{{\Ngal}}{$N_{\rm{H}}^G$}
\newcommand{{\NL}}{$N_{\rm{H}}^L$}

\begin{document}
\SetRunningHead{Author(s) in page-head}{Running Head}
\Received{2000/12/31}
\Accepted{2001/01/01}

\title{Suzaku Observations of the Supernova Remnant N23 in
the Large Magellanic Cloud}

\author{Kentaro \textsc{Someya}$^1$\thanks{Department of Physics, Tokyo Institute of Technology, 2-12-1 Ookayama, Meguro-ku, Tokyo 152-8551, Japan},
Aya \textsc{Bamba}$^{2,1}$, and Manabu \textsc{Ishida}$^1$}
\affil{$^1$Institute of Space and Astronautical Science/JAXA, 3-1-1
  Yoshinodai, Chuo-ku, Sagamihara, Kanagawa 252-5210, Japan}
\affil{$^2$School of Cosmic Physics, Dublin Institute for Advanced Studies,
31 Fitzwilliam Place, Dublin 2, Ireland}
\email{someya@astro.isas.jaxa.jp}


%

\KeyWords{ISM: individual (N23) --- shock waves --- supernova
remnants --- X-ray: ISM} 
\maketitle

\begin{abstract}

 X-ray emission from the supernova remnant N23 in the Large Magellanic
 Cloud (LMC) is studied using the X-ray Imaging Spectrometer (XIS)
 onboard {\suz}. Thanks to superior energy resolution of the XIS in the
 soft X-ray band, we resolved H-like and He-like Oxygen {\ka} emission
 lines from N23 with unprecedentedly high quality, and as a result,
 identified a new optically thin thermal emission component with a
 temperature $\sim$0.2~keV, as well as that with a temperature of
 $\sim$0.5--0.7~keV previously known. This alters the estimate of the
 ionization timescale $n_{\rm e}t$ from $\sim$$10^{10-11}$ cm$^{-3}$~s to
 $\gtrsim$$10^{12}$ cm$^{-3}$~s. Under the assumption that N23 is still
 in the Sedov phase, its age evaluated from the newly discovered low
 temperature component is $\sim$8000~yr, although it is possible that
 N23 has already moved into the radiative phase. The abundances of the heavy
 elements are found to be roughly consistent with those of the LMC
 average, which indicates that the origin of the X-ray emission of N23
 is swept-up ambient material, as expected from its ionization
 timescale.

\end{abstract}

\section{Introduction}

Chemical evolution of the universe has been one of the major issues in
the modern astronomy. Heavy elements, or metals, have been generated and
accumulated in the universe since its birth mainly through
supernovae. An important clue to understand the chemical evolution of
the universe will no doubt be brought about by systematic study of a
well-defined sample of supernova remnants. In X-ray observations, we are
able to extract information of plasma temperature, metal abundance,
explosion energy, ionization age, and so on from optically thin thermal
plasma emission from supernova remnants (SNRs). These pieces of
information will lead to understand galactic chemical evolution and star
formation history. A systematic study of SNRs in the Large Magellanic
Cloud (LMC) suits well for this purpose, because of its well-known
distance (50kpc; \cite{1999PASP..111..775F}) and small interstellar
absorption to LMC. The fact that LMC is face-on to us is another
advantage. Because of these characteristics, systematic X-ray studies of
the thermal SNRs in LMC have been carried out with some major X-ray
observatories (e.g. \cite{1998ApJ...505..732H}). These observations are,
however, limited in terms of energy resolution in the low energy band
below $\sim$1~keV, which is important to study the nature of the SNR
plasma in detail, because the temperature of the plasma in most SNRs is
less than $\sim$2~keV. In particular, hydrogenic and He-like Oxygen
{\ka} lines that appear in 0.5--0.7~keV dominate a spectrum of the
plasma with a temperature of $\lesssim$1~keV, and hence, high resolution
spectroscopy in this energy band is of great importance in evaluating
the parameters of the low temperature plasma, such as the temperature,
the density, the ionization parameter and the abundances of the metal.
%
%
%
%
We therefore have decided to perform a systematic study of the SNRs in
LMC with the Suzaku XIS (\cite{2007PASJ...59S...1M},
\cite{2007PASJ...59S..23K}), which has the best energy resolution in the
Oxygen 0.5--0.7~keV energy band among the CCDs currently in orbit. In
this paper, we study the SNR N23, as a beginning of our systematic study
project.

N23 is identified as a SNR with radio observations at 5 and 14.7~GHz for
the first time (\cite{1980MNRAS.191..469M}). Its X-ray emission extends
$100''\times120''$ in the sky ($24\times29$\,pc at a distance of
50~kpc), showing a semi-circular morphology with only the south-eastern
hemisphere being bright \citep{2006ApJ...645L.117H}. Based on the {\asc}
observation, \citet{1998ApJ...505..732H} classified N23 as a young SNR
with an age of $\sim3800$ from the spectrum of the entire
remnant. \citet{2006ApJ...645L.117H} carried out spatially resolved
spectroscopy of N23 with {\chan}, and estimated the age of north-western
rim to be $\sim$4600~yr. These observations result in consistent
estimation of the temperature and the ionization parameter
($n_{\rm{e}}t$) in the bright south-eastern shell, which is
$\sim$0.5--0.7~keV and $\sim10^{10}$--$10^{11}~\rm{cm^{-3}\,s}$,
respectively. They are, however, not be able to resolve hydrogenic and
He-like {\ka} lines from Oxygen clearly because of limited energy
resolution below $\sim$1~keV. 
In addition, \citet{2006ApJ...645L.117H}
have revealed that there is a point source around the center of N23. From
its 0.5--10~keV flux ($\sim10^{-14}~\rm{erg~cm^{-2}~s^{-1}}$) and
power-law spectrum with a photon index of 2.2, the source is probably a
rotation-powered pulsar and/or a pulsar wind nebula.  If so, the
progenitor of N23 is a core-collapsed massive star
(\cite{2006ApJ...653..280H}).

This paper is organized as follows. After describing how the observation
and data reduction are carried out in \S~\ref{sec:obs}, we report the
results of spectral analysis of N23 in \S~\ref{sec:spec_ana}. Owing to
high spectral resolution of the XIS in the Oxygen {\ka} band, we have
clearly resolved hydrogenic and He-like K$\alpha$ emission lines from
Oxygen for the first time, which leads to identification of a new low
temperature component. Based on these results, we discuss the plasma
parameters and evolutionary phase of N23 in \S~\ref{sec:dis}. Finally,
the summary is given in \S~\ref{sec:sum}.

\section{Observation and Data Reduction}\label{sec:obs}

\begin{figure}[htb]
 \begin{center}
  \FigureFile(80mm,80mm)
  {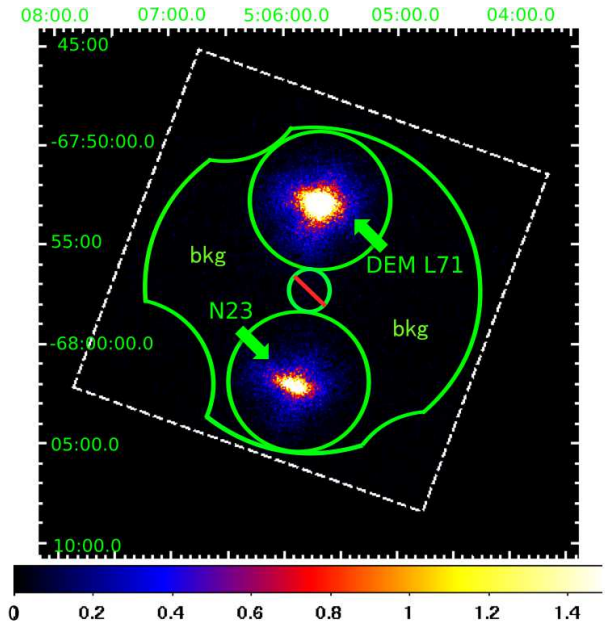}
 \end{center}
 \caption{Suzaku XIS X-ray image of the N23 in the
 bands 0.2--2~keV. Another nearby SNR, DEM L71, is also shown.
 The images from all the
 four XIS modules are combined. The green circles with a radius of
 \timeform{3.'3} are integration regions of the source photons from N23
 .%
 The background photons were extracted from annular region which was
 removed the source and a center regions.
 The calibration sources at the corners are masked. }
 \label{fig:suzaku-image}
\end{figure}

N23 was observed with {\suz} \citep{2007PASJ...59S...1M} on
2005 August 16--17. The observation log is summarized in
table~\ref{tab:suzakuobslog}. 
 \begin{table}[tbh]
  \begin{center}
    \caption{observation log.}
   \label{tab:suzakuobslog}
    \begin{tabular}{lll}\hline
     Observation ID   & \multicolumn{2}{l}{100003010} \\
     Observation mode & \multicolumn{2}{l}{Full window, no burst, no SCI} \\
     Start Date (UT)  & \multicolumn{2}{l}{2005/08/16--17} \\
     Exposure   (ksec)& \multicolumn{2}{l}{7.1} \\
     Count rate$^*$ (cts~s$^{-1}$) & FI: 0.6 & BI: 1.2 \\\hline
    \end{tabular}
  \end{center}
  {$^*$ Count rate of FI and BI-CCD. Count rate of N23 in the
  0.5--2.0 band within a $3'.3$ radius circle after
  the background subtraction.}
 \end{table}
{\suz} is equipped with four modules of the XIS
(\cite{2007PASJ...59S..23K}). One of
them adopts a back-illuminated CCD (BI-CCD), which is referred to as
XIS1 having superior quantum efficiency especially in a low energy band,
whilst the other three modules (XIS0, XIS2 and XIS3) utilize
front-illuminated (FI) CCDs which have high sensitivity in a high energy
band and better energy resolution than the BI-CCD. They are adapted at
the focal plane of the X-Ray Telescopes (XRT;
\cite{2007PASJ...59S...9S}). They are thin-foil-nested X-ray mirrors
realizing a high throughput in 0.2--12~keV with a moderate imaging
capability (\timeform{1.'8}--\timeform{2.'3} in half-power diameter).
The {\suz} XIS image of N23 below 2~keV is shown in Fig.~\ref{fig:suzaku-image}.

Data reduction and analysis of the present data were carried out using
the HEADAS software package version 6.5.1. The data processed by {\suz}
pipeline processing software (ver 2.0) were analyzed. 
We only used the data taken with the $5\times5$ editing mode
in the high and super-high data rates because of the editing mode of
$2\times2$ has uncertainty in the response matrices of the XIS.
In screening the
data, we removed time intervals while the elevation angle from the night
earth is less than 5$^\circ$. 
In the standard data screening procedure, the time interval during which
the elevation angle from the day earth is less than 20$^\circ$. This
results in, however, only 5.6~ksec data available. We thus reduce the
day-earth elevation angle by degrees by closely comparing resulting
spectra and have finally found that the solar X-ray contamination does
not raise any spectral modification even if we reduce the day earth
elevation down to 5$^\circ$ (DYE\_ELV $>5^\circ$) for the XIS
data. Owing to this study, the total exposure time has increased to
7.1~ksec.

\section{Spectral Analysis}\label{sec:spec_ana}

\subsection{Extraction of Source and Background Photons}

In extracting source photons, we adopt a circular aperture with a radius
of \timeform{3.'3} centered on N23 (the green circle centered on N23 in
Fig.~\ref{fig:suzaku-image}). For the background, we have collected
photons from the entire CCD area out of the source extraction region,
except for outer edge regions illuminated by $^{55}$Mn isotopes adapted
in the camera bodies of all the XIS modules, the central circular region
with a radius of $\sim$\timeform{1.'5}, and the \timeform{3.'3} circular
region centered on the other SNR {\dem}. We combined the spectra from
the XIS adopting the FI-CCDs (XIS0, XIS2, and XIS3) into a single
spectrum, and refer to it as the FI spectrum hereafter. The BI spectrum
is the same as that solely from the XIS1 data. 

In the spectral analysis, we use {\sc xspec} version 11.3.2aj.
The XIS response matrix (RMF) and auxiliary response file (ARF)
are calculated using {\sc xisrmfgen} (version 2007-05-14) and {\sc
xisarfgen} (version 2008-04-05), respectively
(\cite{2007PASJ...59S.113I}). The RMFs and ARFs from
the FI-CCDs are combined with the ftools {\sc addrmf} and {\sc marfrmf}.

\subsection{Spectral Models}

From the {\asc} observations of LMC SNRs, it is found that the spectra
of N23 is well represented by an optically thin thermal
emission model in ionization non-equilibrium
\citep{1998ApJ...505..732H}. 
In characterizing the spectra of SNR, we thus adopt the
non-equilibrium ionization model, or {\sc vnei} model in the {\sc xspec}
model library. The {\sc vnei} model provides the plasma temperature
$kT_{\rm e}$, the ionization timescale $n_{\rm e}t$ where $t$ is the
elapsed time since the shock occurs, the metal abundances, and the
normalization parameter $\frac{10^{-14}}{4\pi D^2}\int n_{\rm e}n_{\rm
H}dV$ where $D$ is the distance to the source, $n_{\rm e}$ and $n_{\rm
H}$ are the electron and hydrogen density respectively, and $dV$ is the
volume element of the SNR.

The emission spectra are attenuated by photoelectric absorption due to
metals contained in the interstellar matter within our galaxy and
LMC. Since the metal abundances of these two absorption components are
significantly different, we consider them separately in the spectral
evaluation as follows. We assume solar composition
\citep{1989GeCoA..53..197A} for the absorbing matter in the Galaxy, and
denote its hydrogen column density as {\Ngal}. The value of {\Ngal}
toward N23 is obtained by the Galactic H{\footnotesize I} survey, which
is 5.8$\times10^{20}$cm$^{-2}$ \citep{1990ARA&A..28..215D}. We utilize
the {\sc phabs} model in {\sc xspec} to represent the galactic
absorption, and freeze {\Ngal} at these values in the spectral fitting
described below. The average metal abundances of LMC, on the other hand,
are measured by \citet{1992ApJ...384..508R}, which is $\sim$0.3 solar on
average over major metals. We reflect these abundances on the {\sc
vphabs} model, and set the column density associated with LMC {\NL} free
to vary in the following spectral fit process.

\begin{longtable}{ccccc}
	\caption{Best-fit parameters for N23\footnotemark[$*$].}\label{tab:N23bestfitpar}
	\hline
	  Parameters & $1kT1nt$ & $1kT2nt$ & $2kT1nt$ & $2kT2nt$ \\
	\hline
	\endhead
	\hline
	\endfoot
	\hline
    $N_{\rm{H}}^{G}$\,($\times10^{20}\,\rm{cm^{-2}}$) & 5.8(fix) &
     5.8 (fix) & 5.8(fix) & 5.8(fix) \\
    $N_{\rm{H}}^{L}$\,($\times10^{21}\,\rm{cm^{-2}}$) & 1.6(1.3--1.9) &
     $<0.16$ &
     1.5(0.6--1.9) & 0.9(0.6--1.9)\\
    $kT_{\rm{e}1}$\,(keV) & 0.548(0.544--0.567) & 0.572(0.557--0.580) &
     0.569(0.556--0.593) & 0.581(0.551--0.597) \\
    $kT_{\rm{e}2}$\,(keV) & -- & -- & 0.218(0.214--0.230) & 0.225(0.214--0.230)\\
    $\log (n_et)_1$\,($\rm{cm^{-3}\,s }$) & 10.51(10.48--10.55) &
     10.50(10.45--10.52)  & 13.68(11.92--13.70) & 12.04(11.87--13.70) \\
    $\log (n_et)_2$\,($\rm{cm^{-3}\,s}$) & -- & 13.18(12.77--13.70) &
     --  & 13.63(13.27--13.70)\\ 
    $Norm_1$\footnotemark[$\dagger$] ($\times10^{-2}$) &
     1.87(1.74--1.90) & 0.34(0.31--0.68) & 0.83(0.64--1.00)
     & 0.78(0.66--0.98)\\
    $Norm_2$\footnotemark[$\dagger$] ($\times10^{-2}$) & -- &
     0.73(0.67--0.81) & 2.54(1.93--2.96) & 1.93(1.47--3.51)\\
     C & 0.30(fix) & 0.30(fix) & 0.30(fix) & 0.30(fix)\\
     N & $<0.02$ & 0.09(0.04--0.25) & 0.54(0.12--1.05) &
     0.37(0.23--0.94)\\
     O & 0.093(0.089--0.098)  & 0.27(0.26--0.28) & 0.33(0.25--0.50) &
     0.34(0.28--0.49) \\
     Ne & 0.15(0.14--0.16) & 0.43(0.39--0.46) & 0.49(0.35--0.70) &
     0.49(0.40--0.66)\\
     Mg & 0.14(0.12--0.17) & 0.31(0.27--0.35) & 0.39(0.33--0.60) &
     0.39(0.32--0.47)\\
     Si & 0.28(0.19--0.36) & 0.34(0.24--0.43) & 0.41(0.30--0.52) &
     0.42(0.32--0.55) \\
     Fe & 0.111(0.105--0.117) & 0.18(0.17--0.22) & 0.24(0.19--0.33) &
     0.24(0.19--0.33) \\
    $\chi^2_{\nu}$(d.o.f.) & 1.45(99) & 1.11(97) & 1.00(97) & 1.00(96)\\
    offset FI (eV) & $-$8.7($-$11.0--$-$7.8) & $-$8.1($-$9.1--$-$5.9) &
     $-$8.1($-$9.0--$-$7.1) & $-$8.0($-$9.0--$-$7.1)\\
    offset BI (eV) & $-$1.0($-$1.6-- 0.0)   & $-$1.0($-$2.1-- 0.0) &
     -1.0($-$2.8-- 0.0) & $-$0.3($-$1.3-- 0.0) \\\hline
     \multicolumn{3}{l}{\hbox to 0pt{\parbox{180mm}{\footnotesize
     \footnotemark[$*$] The errors in the parentheses represent the 90\% confidence intervals.\\
     \footnotemark[$\dagger$] In units of $10^{-14}/4\pi D^2\int n_{\rm{e}}n_{\rm{H}}dV\,\rm{cm^{-5}}$, 
     where $V$ and $D$ are the Volume and distance to the plasma, respectively.
      }}}
      \endlastfoot
\end{longtable}
In evaluating the spectrum of N23 through spectral fitting, we begin
with a single component {\sc vnei} model attenuated by photoelectric
absorption. Since the carbon abundance cannot be constrained because of
the large interstellar absorption at carbon {\ka} line energies, we
fixed it to the LMC value \citep{1992ApJ...384..508R}. 
In the early phase of the Suzaku mission, the energy
gain has a large uncertainty \citep{2007PASJ...59S..23K}. The gain
uncertainty is $\sim\pm$10~eV, according to the previous studies on SNRs
\citep{2008PASJ...60S.153B,2008PASJ...60S.141Y}. We therefore have
allowed the energy offset to be floated. As a result, the energy offset
is converged to $\sim -$8~eV and $\sim -$1~eV, for the FI and BI
spectra, respectively. These gain offset values are within the gain
uncertainty. The result is summarized in the second column of
table~\ref{tab:N23bestfitpar} labelled `$1kT1nt$'. The best fit results
in $kT_{\rm{e}}\simeq0.548$~keV and
$\log(n_{\rm{e}}t~[\rm{cm^{-3}~s}])\simeq10.51$ with a $\chi^2$ (d.o.f.)
of 144 (99). These parameters are roughly consistent with those from the
{\chan} observation. Note that, if we do not float the energy gain, the
best-fit parameters are $kT_{\rm{e}}\simeq0.552$~keV and
$\log(n_{\rm{e}}t~[\rm{cm^{-3}~s}])\simeq10.54$ with a $\chi^2$ (d.o.f.)
of 219 (101). Hence, the energy offset adjustment significantly improves
the fit, while the differences in the temperature and the ionization
timescale (in logarithm) are within $\sim1$~\%.

\begin{figure}
 \begin{center}
  \FigureFile(80mm,80mm)
  {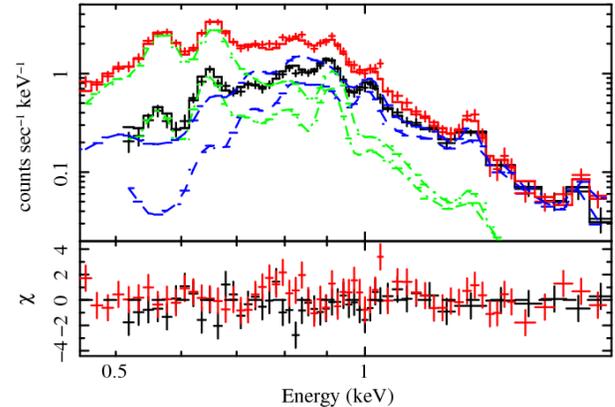}
 \end{center}
 \caption{The data and the best-fit `$2kT1nt$' model of N23. in the
 0.45--2.0~keV band. The black and red crosses show the data point from
 the FI and BI CCDs, respectively. The solid histograms with the same
 color are the best-fit model. The blue and green broken histograms show
 the components with $kT_{\rm{e}}$=0.57 and 0.22~keV, respectively. The lower
 panel shows the residuals from the best-fit model.
 \label{fig:xis_N23_2kt1nt}}
\end{figure}

As noticed from this table, however, the single component {\sc vnei}
model no longer provides an acceptable fit ($\chi^2_{\nu} \simeq$ 1.5).
We thus have appended another {\sc vnei} model. To reduce the number of
model parameters as much as possible, we constrained either $kT_{\rm e}$
or $n_{\rm e}t$ common between the two {\sc vnei} components, and fit
them to the data separately. The resultant best-fit parameters are also
listed in table~\ref{tab:N23bestfitpar}, labelled `$1kT2nt$' and
`$2kT1nt$', respectively. Although the $1kT2nt$ model improves the fit
significantly, the reduced $\chi^2$ value indicates that it is still not
acceptable. The improvement of the $2kT1nt$ model, on the other hand, is
more remarkable, and it provides an acceptable fit at the 90\%
confidence level. The best-fit model is shown in
Fig.~\ref{fig:xis_N23_2kt1nt}.
This result indicates that there exists another optically thin thermal
plasma component with a temperature of $\sim$0.2~keV, in addition to the
0.5--0.6~keV component so far known.
The discovery of the new low temperature component is brought about by
the clear resolution of the Oxygen {\ka} lines by the {\suz} XIS.
Consequently, the best-fit value of the ionization timescale now becomes
$\log (n_{\rm e}t\,~\rm{[cm^{-3}\,s]}) \gtsim 12$. This is significantly
larger than that from the {\asc} and {\chan} observations
($10^{10}-10^{11}~\rm{cm^{-3}~s}$).

In the {\chan} observation, more than $\sim30$~\% of the
total emission comes from the eastern rim
\citep{2006ApJ...645L.117H}. Since our {\suz} observation covers the
remnant entirely, the best-fit parameters we have obtained are a kind of
mean weighted mainly on this extended bright rim region (see Fig.~\ref{fig:N23image}).

We finally remove the constraint of the ionization timescale between the
two {\sc vnei} components and set them free to vary independently. The
results are summarized in the `$2kT2nt$' column of
table~\ref{tab:N23bestfitpar}. This change, however, is not effective in
improving the fit. We therefore adopt the $2kT1nt$ model for N23 for
discussion in later sections.

\section{Discussion}\label{sec:dis}

\subsection{Densities, total energy, and swept up mass}\label{sec:N23_phys}

In this section, we aim to derive electron number densities of the SNR
plasma ($n_{\rm e}$) and ambient ISM ($n_0$), total thermal energy of
the plasma ($E_{\rm t}$), swept up mass ($M_{\rm swept}$). With the aid of
the normalization obtained from the spectral fitting, the volume of the
plasma ($V$) and the volume filling factor ($f$), we calculated
$n_{\rm{e}}$, taking into account $n_{\rm{e}}$=1.2$n_{\rm{H}}$ for fully
ionized solar abundance plasma.  Assuming the strong shock, pre-shock
density (e.g. ISM density; $n_0$) is obtained to be
$n_0 = (n_{\rm H} / 4)$.
The swept-up mass is given by $M_{\rm swept} = n_{\rm{H}}m_{\rm p}fV$,
where $m_{\rm{p}}$ is the proton mass.
The total energy is given by $E_{\rm t}=3n_{\rm e}kTfV$, where $k$ and
$T$ is the Boltzmann constant and the temperature of the plasma,
respectively. Note that we assume that electrons and ions temperature
are in thermal equilibrium.
The age of the plasma $t_{\rm ion}$ is obtained from the ionization
parameter ($n_{\rm{e}}t$) which is evaluated from the
spectral fitting.

\begin{figure}
 \begin{center}
  \FigureFile(80mm,80mm)
  {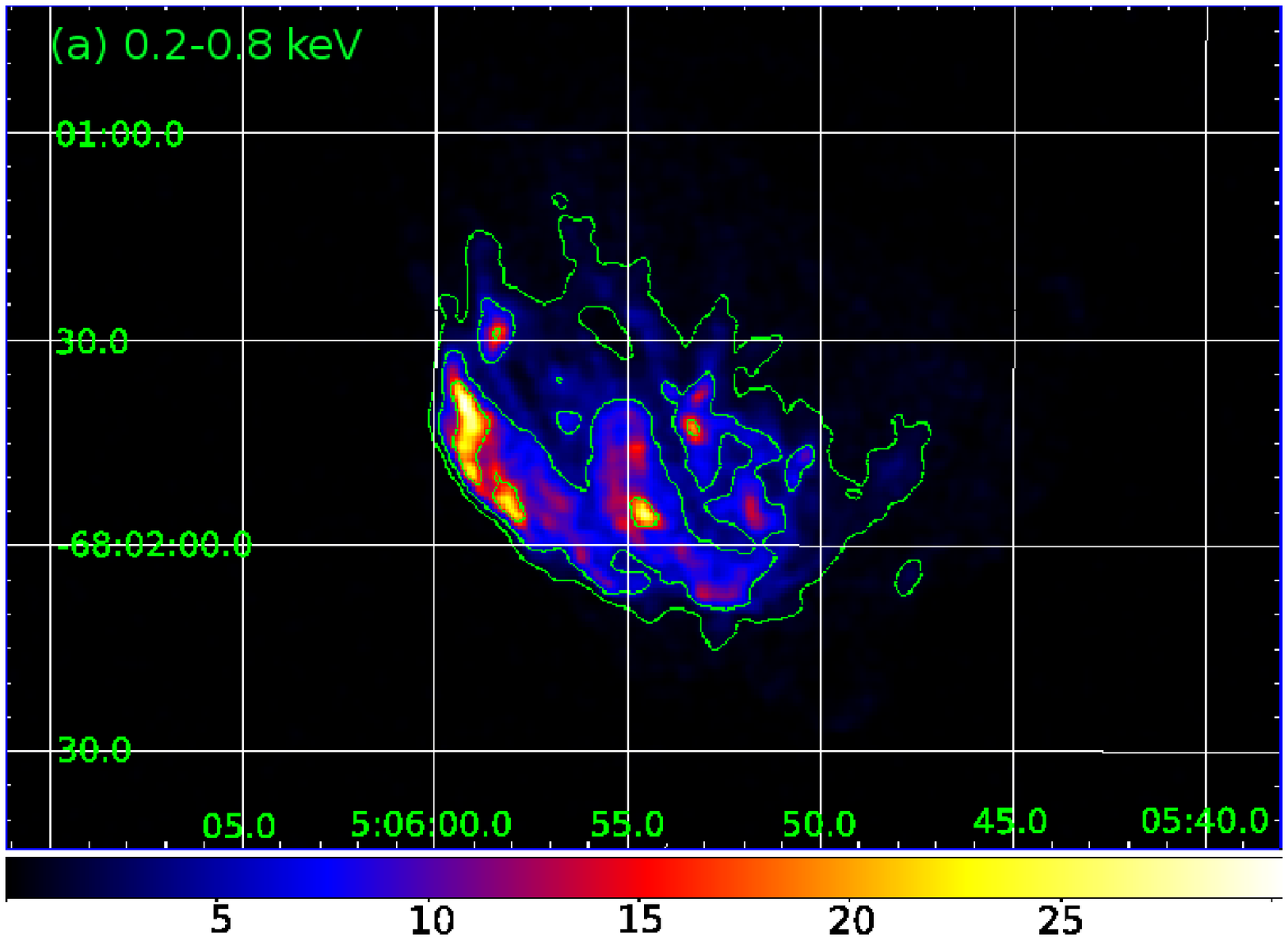}
  \FigureFile(80mm,80mm)
  {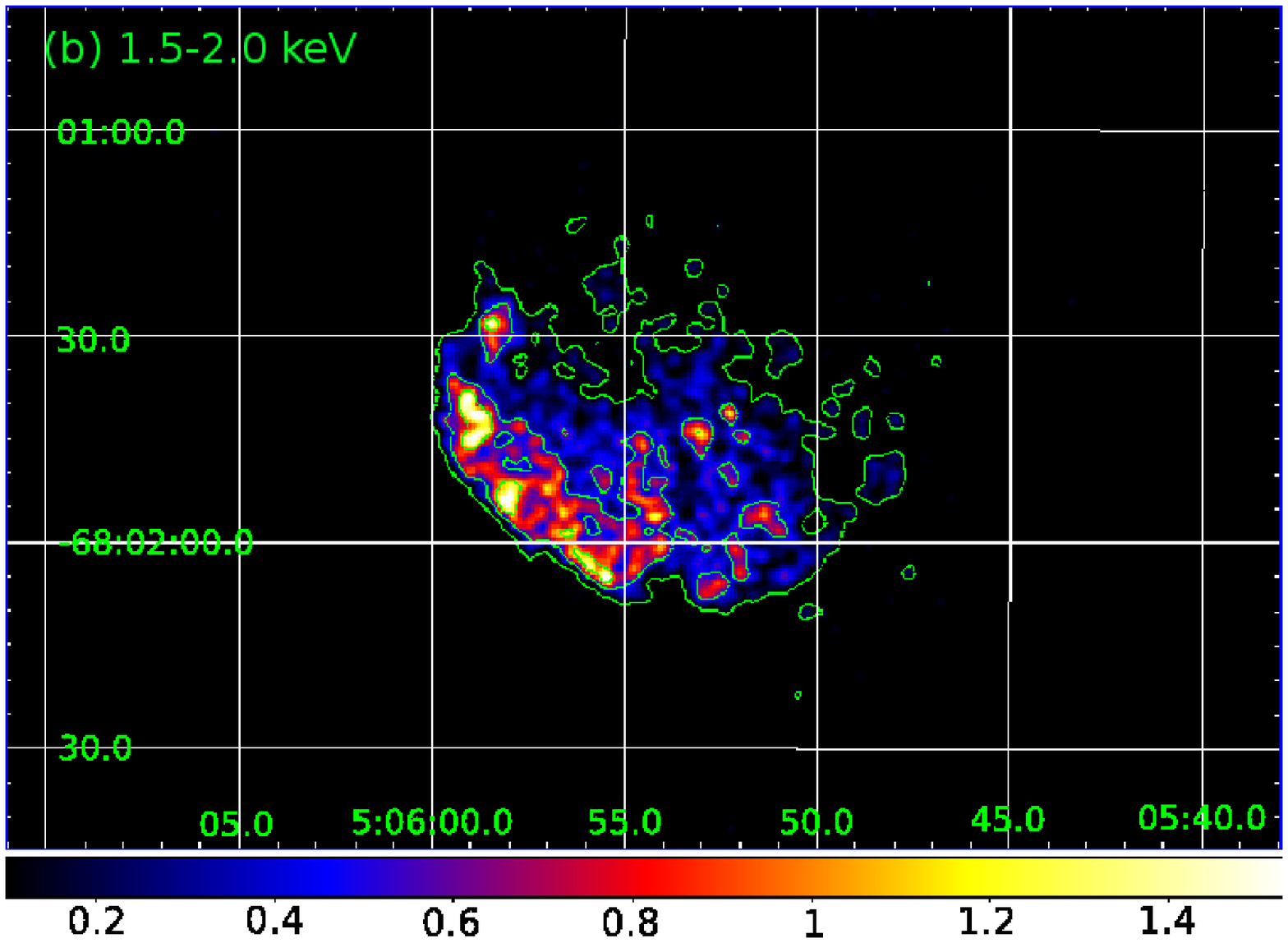}
 \end{center}
 \caption{Smoothed {{\chan}} image of N23 (a) in the 0.2--0.8~keV and
 (b) in the 1.5--2.0~keV band. The contours overlaid with labels of
 1.0, 5.0, 15.0 (left) and 0.1, 0.5, 1.0 counts
 pixel$^{-1}$}\label{fig:N23image}
\end{figure}

In order to estimate the densities, we need to know the volume of the
plasma from an X-ray image. Since the {\suz} XIS has only moderate
imaging capability, we carried out {\chan} imaging analysis first.
%
%
We analyzed the {\chan} ACIS data of N23 (seq\# 500212) observed on 2002
December 29--30.
Images are extracted from a standard cleaned event file using the {\sc
dmcopy} command. Fig.~\ref{fig:N23image} shows the X-ray images in the
(a) 0.2--0.8~keV and (b) 1.5--2.0~keV energy bands.
The low and high energy bands were chosen to show the contribution from
the $\sim$0.2~keV and the $\sim$0.6~keV components, respectively.
%
In both images, the structure of N23 shows a bright hemisphere with
$\sim$35$''$ (8.5~pc) radius.  We therefore assumed that the N23 plasma
distributes in the shell of the hemisphere with a radius and a thickness
of $35''$ and $2.7''$ (0.7~pc), the latter of which is 1/12 of the shell
radius expected for the ISM compressed in the case of the strong and
adiabatic shock, which assumption is to be considered later in
\S~4.2. As a result, the volume of the plasma becomes $V =
9.3\times10^{57}$ cm$^3$.

The other parameters are summarized in table~\ref{tab:N23phi}. 
\begin{table}
 \caption{Input and derived properties for N23.}
 \label{tab:N23phi}
 \begin{center}
  \begin{tabular}{lcc}\hline
   Property & \multicolumn{2}{c}{Value} \\\hline
   & low-$T$ component & high-$T$ component \\
   $kT_{\rm{e}}$ (keV)           & 0.22 & 0.57 \\
   $R$ (pc)               & 8.5 & 8.5 \\
   $V$ ($10^{57}$~cm$^{3}$) & 9.3 & 9.3 \\
   $n_{\rm{e}}$ (cm$^{-3}$)           & 
   31.3$^{+2.5}_{-4.1}f_{0.1}^{-1/2}$ &
   18.0$^{+1.6}_{-2.2}f_{0.1}^{-1/2}$ \\
   $n_0$ (cm$^{-3}$)           & 
   7.9$^{+0.6}_{-1.0}f_{0.1}^{-1/2}$ & 
   4.4$^{+0.3}_{-0.6}f_{0.1}^{-1/2}$ \\
   $E_{\rm{t}}$ ($10^{50}$~erg) & 
   0.3$\pm0.1f_{0.1}^{1/2}$ & 0.4$\pm0.1f_{0.1}^{1/2}$ \\
   $M_{\rm{swept}}$ (M$_\odot$)       & 
   21$^{+2}_{-3}f_{0.1}^{1/2}$ & 
   12$^{+1}_{-2}f_{0.1}^{1/2}$ \\
   $t_{\rm{ion}}$ (yr) & $>1000f_{0.1}^{1/2}$ & $>1500f_{0.1}^{1/2}$ \\\hline
  \end{tabular}
 \end{center} 
\end{table}
As shown in Fig.~\ref{fig:N23image}, the emission from N23 is highly
anisotropic.  We therefore normalized the volume filling factor $f$ by
0.1 ($\equiv f_{0.1}$). 
The plasma density estimated from the low and high
temperature components are $\sim31~\rm{cm^{-3}}$ and
$\sim18~\rm{cm^{-3}}$, respectively, as shown in table
\ref{tab:N23phi}. This density estimation is roughly consistent with
that of the southeastern rim region with the {\chan} observation
($\sim10$--$23~\rm{cm^{-3}}$; \cite{2006ApJ...645L.117H}).
With this normalization, the ISM density estimated from the low and high
temperature components are $\sim8~\rm{cm^{-3}}$ and
$\sim4~\rm{cm^{-3}}$, respectively.  One may imagine that this high
density is realized through collision of the ejecta with molecular
clouds. According to a radio observation \citep{1997ApJ...480..607B},
however, no CO emission was discovered from a region including N23. We
thus consider that N23 explodes into a relatively dense part of the ISM
($\sim10~\rm{cm^{-3}}$).
It is suggested this high density is related
geometrically with the open cluster HS 114
\citep{1966AJ.....71..363H} near the remnant
\citep{2006ApJ...645L.117H}.
Finally, total ejecta mass is given by $M_{\rm{low}}+M_{\rm{high}}$,
where $M_{\rm{low}}$ and $M_{\rm{high}}$ are the mass of low and high
temperature components, respectively.  Total ejecta mass therefore is
calculated to be $\sim$30$~M_\odot$. Hence the plasma chiefly comprises
of the swept-up ISM.

\subsection{SNR phase}\label{sec:phase}

In this section, we consider evolutionary phase of N23 by comparing
various timescales evaluated from observed parameters. We compare the
Sedov time scale ($t_{\rm{Sedov}}$), the plasma age ($t_{\rm ion}$)
estimated from the ionization parameter, and the cooling time scale
($t_{\rm cool}$) of N23. $t_{\rm Sedov}$ is based on the concept of
Sedov simple blast wave model (\cite{1959sdmm.book.....S}), in which a
supernova with explosion energy ($E_0$) expands into homogeneous ISM,
which is given by
\begin{equation}
\label{eq:sedovage}
 t_{\rm Sedov}=4.3\times10^2\left(\frac{R}{1~\rm{pc}}\right)
\left(\frac{kT_e}{1~\rm{keV}}\right)^{-0.5}~\rm{yr},
\end{equation}
where $R$ is the blast wave shock radius and $T_{\rm{e}}$ is the shock
temperature. $t_{\rm cool}$ is defined as the time scale of the
temperature distribution having started to deviate from that of the
Sedov model, which is written as
\begin{equation}
\label{eq:coolage}
 t_{\rm cool}=2.7\times10^4\left(\frac{E_0}{10^{51}~\rm{erg}}\right)^{0.24}
\left(\frac{n_0}{1~\rm{cm^{-3}}}\right)^{-0.52}~\rm{yr}
\end{equation}
\citep{1981MNRAS.195.1011F}. If $t_{\rm cool}\simeq t_{\rm Sedov}$
holds, the SNR is considered to locate at a late stage of the Sedov
phase or at an initial radiative phase. If, on the other hand,
$t_{\rm{Sedov}} < t_{\rm{cool}}$ holds, the SNR is regarded as being
still in the Sedov phase. We assume that the explosion energy is
$10^{51}$~erg in eq.~(\ref{eq:coolage}).

We discovered the new soft emission component with a
temperature of $\sim$0.22~keV, in addition to the 0.5--0.7~keV component
so far known. Accordingly, we need to update the remnant age estimation.
The Sedov time scale and the cooling time scale calculated with
eq.~(\ref{eq:sedovage}) and (\ref{eq:coolage}) are summarized in
table~\ref{table:ages}, together with the ionization age $t_{\rm ion}$ 
obtained based on the spectral fit.
Since the spectral fit with the `$2kT1nt$' model
provides only the lower limit for the ionization timescale $n_{\rm e}t$,
we only have a lower limit for $t_{\rm{ion}}$. We thus have decided to
estimate the remnant age with eq.~(1) and (2).
\begin{table}
  \caption{The characteristic ages for N23.}\label{table:ages}
  \begin{center}
   \begin{tabular}{lcccccc}\hline
    & low-$T$ component & high-$T$ component\\\hline
    $t_{\rm{Sedov}}$ ($10^4~\rm{yr}$) & 0.8 & 0.5 \\
    $t_{\rm{cool}}$  ($10^4~\rm{yr}$) & 0.9$f_{0.1}^{0.26}$ &
    1.2$f_{0.1}^{0.26}$ \\
    $t_{\rm{ion}}$ ($10^4~\rm{yr}$)   & $>0.1f_{0.1}^{1/2}$ &
    $>0.2f_{0.1}^{1/2}$ \\\hline
   \end{tabular}
  \end{center}
\end{table}
Using eq.~(\ref{eq:sedovage}) with $R = 8.5$~pc,
$t_{\rm{Sedov}}$ estimated from the low and high temperature components
becomes $\sim$8000 and $\sim$5000~yr, respectively, which are consistent
with the lower limit of $t_{\rm{ion}}$. This is a direct consequence of
eq.~(\ref{eq:sedovage}), implying $t_{\rm{Sedov}}$ being proportional to
$T_{\rm{e}}^{-1/2}$. The high temperature component is younger than the
low temperature component. In the Sedov scheme, this can be interpreted
as the high temperature component being heated recently.
%
On the other hands, $t_{\rm cool}$ of the low and high temperature
components are 9300$f_{0.1}^{0.26}$ and 12000$f_{0.1}^{0.26}$,
respectively, using eq.~(\ref{eq:coolage}) with $n_0$ given in
table~\ref{tab:N23phi} (\S\ref{sec:N23_phys}). Since $t_{\rm Sedov}
\lesssim t_{\rm cool}$ for both components, N23 is likely to be at a
late stage of the Sedov phase, and if so, its age is $\sim$8000~yrs old
from the newly discovered low temperature component.

Note, however, that this discussion depends on the
estimation of the volume filling factor $f$. Accounting for the highly
anisotropic image (Fig.~\ref{fig:N23image}), we have normalized $f$
by 0.1. However, if the shock wave of N23 propagates in a higher density
ISM, like the edge of the open cluster HS~114
\citep{1966AJ.....71..363H}, the total volume should be smaller for the
given emission measure, resulting in a smaller $f$. If, for instance, $f
= 0.01$, $t_{\rm{cool}}$ of the low and high temperature components are
$\sim$5000 and $\sim$7000~yr, respectively. Since $t_{\rm{cool}}
\lesssim t_{\rm{Sedov}}$, the low temperature component should be
regarded as in the radiative phase. In addition, we would also like to
remark that, in \S~\ref{sec:N23_phys}, we have calculated the plasma
volume, assuming that the shell thickness is 1/12 of the shell radius,
which is expected for a spherical strong shock. This holds only if N23
is still in the Sedov (adiabatic) phase. If N23 has been in the
radiative phase, then the shell thickness should be smaller due to
radiative cooling, and hence, the plasma density would be larger than
the current estimation. As a result, the cooling time scale estimated
from eq.~(\ref{eq:coolage}) should be smaller, and $t_{\rm cool}
\lesssim t_{\rm Sedov}$. Consequently, we cannot deny possibility that
N23 has been in the radiative phase.

In summary, the age of N23 is estimated to be $\sim$8000
yr from the newly discovered low temperature component as long as it
still stays in Sedov phase. This condition is, however, uncertain,
because the discussion depends on the volume filling factor of the
plasma. If, for instance, the shock propagate in the higher density ISM,
N23 may have already entered into the radiative cooling phase. In this
case, the remnant age estimated from the cooling timescale is smaller.

Finally, we can make an independent age estimation from
the central source detected by {\chan} \citep{2006ApJ...653..280H}. Its
luminosity is $1\times10^{34}$~erg in the 0.5--10.0~keV
\citep{2006ApJ...653..280H}. If the central source is a pulsar (although
the pulsation has not been detected), the characteristic age is
$\sim10^3$--$10^{4.5}$~years \citep{2002A&A...387..993P}, which is
consistent with our age estimation update.

\subsection{Abundance}

\begin{figure}
 \begin{center}
  \FigureFile(80mm,80mm)
  {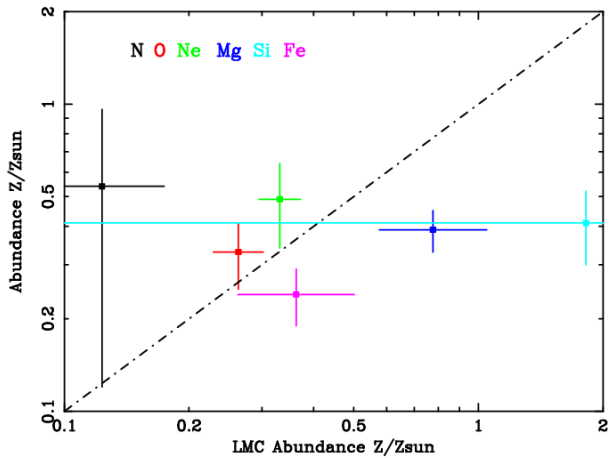}
 \end{center}
 \caption{Elemental abundance as derived from fits to the XIS data. The
 derived N23 abundance value is the square. 
 The ratios are all relative to solar (Andres \& Grevesse
 1989). The silicon abundance in Russell \& Dopita's work is highly 
 uncertain as indicated by our use of a large error bar for this
 species. Dashed line represents the abundance in these SNRs are same to
 the LMC average.}
 \label{fig:ab}
\end{figure}

The main component of X-ray emission originates from the swept-up ISM in
N23. Hence the elemental abundances of N23 reflect those of its
environment. Figure~\ref{fig:ab} shows the abundances obtained from the
spectral analysis (table~\ref{tab:N23bestfitpar}).
The abundances of N, O, Ne and Fe agree roughly with the LMC average
\citep{1992ApJ...384..508R}. The Si and Mg abundances are smaller than
those obtained (\cite{1992ApJ...384..508R}), although their errors are
relatively large. The Mg and Si abundances obtained by Suzaku are
consistent with those from ASCA (\cite{1998ApJ...505..732H}),
respectively, which seem slightly less than the LMC average.

\section{SUMMARY}\label{sec:sum}

We observed N23 in the Large Magellanic Cloud with Suzaku, and
discovered the new soft emission component with a temperature of
$\sim0.22$~keV, in addition to the $\sim0.5$--0.7~keV component so far
known. This alters the estimate of the ionization parameter ($n_{\rm
e}t$) significantly from $\sim10^{10}-10^{11}~\rm{cm^{-3}~s}$ to
$\gtrsim10^{12}~\rm{cm^{-3}~s}$.
With the aid of {\chan} imaging capability, we are able
to calculate the density of the plasma, and have confirmed that the
supernova explosion of N23 occurred in a high density region with an ISM
density of $\sim$4--8~cm$^{-3}$. The relatively high ambient density may
be related to the open cluster HS 114 \citep{1966AJ.....71..363H} near
the remnant \citep{2006ApJ...645L.117H}.
The parameters of the plasma from our analysis indicate that N23 is
either at a late stage of the Sedov phase or in the radiative cooling
phase. Assuming it is still in the Sedov phase, we have estimated the
age of the remnant to be $\sim$8000~yr from the newly discovered soft
component, which is twice as old as the estimation by
\citet{1998ApJ...505..732H}. Given the uncertainty of
the volume filling factor, however, we cannot deny the possibility that
N23 has already entered into the radiative phase.

The abundances of N23 is roughly consistent with the LMC average
(\cite{1992ApJ...384..508R}). The Si and Mg abundances obtained by
spectral fitting are smaller than those obtained by
\citet{1992ApJ...384..508R}, although their errors on Si and Mg
abundances are relatively large. The Mg and Si abundances obtained by
{\suz} are consistent with those from {\asc}
(\cite{1998ApJ...505..732H}).

\bigskip
We thank Dr. Y. Maeda for his useful discussion and comments. We would
like to express our gratitude to all members of the Suzaku team for
their contributions to the instrument preparation, spacecraft operation,
software development, and in-orbit instrumental calibration.
We also thank an anonymous referee for useful comments.








\begin{thebibliography}{}

\bibitem[Anders \& Grevesse(1989)]{1989GeCoA..53..197A} Anders, E., \
\& Grevesse, N.\ 1989, \gca, 53, 197 

\bibitem[Bamba et al.(2008)]{2008PASJ...60S.153B} Bamba, A., et al.\ 2008, 
\pasj, 60, 153 

\bibitem[Banas et al.(1997)]{1997ApJ...480..607B} Banas, K.~R., Hughes, 
J.~P., Bronfman, L., \& Nyman, L.-A.\ 1997, \apj, 480, 607 

\bibitem[Dickey \& Lockman(1990)]{1990ARA&A..28..215D} Dickey, J.~M., \&
Lockman, F.~J.\ 1990, \araa, 28, 215

\bibitem[Falle(1981)]{1981MNRAS.195.1011F} Falle, S.~A.~E.~G.\ 1981, 
\mnras, 195, 1011 

\bibitem[Feast(1999)]{1999PASP..111..775F} Feast, M.\ 1999, \pasp, 111, 775 

\bibitem[Hayato et al.(2006)]{2006ApJ...653..280H} Hayato, A., Bamba,
A., Tamagawa, T., \& Kawabata, K.\ 2006,
\apj, 653, 280 

\bibitem[Hodge 
\& Sexton(1966)]{1966AJ.....71..363H} Hodge, P.~W., \& Sexton, J.~A.\ 1966, \aj, 71, 363 

\bibitem[Hughes et al.(1998)]{1998ApJ...505..732H} Hughes, J.~P., Hayashi, 
I., \& Koyama, K.\ 1998, \apj, 505, 732 

\bibitem[Hughes et al.(2006)]{2006ApJ...645L.117H} Hughes, J.~P., Rafelski, 
M., Warren, J.~S., Rakowski, C., Slane, P., Burrows, D., \& Nousek, J.\
2006, \apjl, 645, L117 

\bibitem[Ishisaki et al.(2007)]{2007PASJ...59S.113I} Ishisaki, Y., et al.\ 
2007, \pasj, 59, 113 

\bibitem[Koyama et al.(2007)]{2007PASJ...59S..23K} Koyama, K., et al.\ 
2007, \pasj, 59, 23 




\bibitem[Milne et al.(1980)]{1980MNRAS.191..469M} Milne, D.~K., Caswell, 
J.~L., \& Haynes, R.~F.\ 1980, \mnras, 191, 469 

\bibitem[Mitsuda et al.(2007)]{2007PASJ...59S...1M} Mitsuda, K., et al.\ 
2007, \pasj, 59, 1 

\bibitem[Possenti et 
al.(2002)]{2002A&A...387..993P} Possenti, A., Cerutti, R., Colpi, M., \& Mereghetti, S.\ 2002, \aap, 387, 993 

\bibitem[Russell \& Dopita(1992)]{1992ApJ...384..508R} Russell, S.~C.,
\&Dopita, M.~A.\ 1992, \apj, 384, 508 

\bibitem[Sedov(1959)]{1959sdmm.book.....S} Sedov, L.~I.\ 1959,
				  Similarity and Dimensional Methods in
				  Mechanics, New York: Academic Press,
				  1959,  

\bibitem[Serlemitsos et al.(2007)]{2007PASJ...59S...9S} Serlemitsos, P.~J., 
et al.\ 2007, \pasj, 59, 9 

\bibitem[Yamaguchi et al.(2008)]{2008PASJ...60S.141Y} Yamaguchi, H., et 
al.\ 2008, \pasj, 60, 141 

\end{thebibliography}
\end{document}